\documentclass[aps, twoside, nofootinbib]{revtex4}
\usepackage{amsfonts}
\usepackage{amssymb}
\usepackage[mathscr]{eucal}
\usepackage[dvips]{graphicx}
\usepackage{amsfonts,amssymb,amsthm}
\usepackage{amsmath}

\def \ni{\noindent}

\def \txt{\textrm}

\def \be{\begin{equation}}
\def \ee{\end{equation}}
\newcommand{\ben}{\begin{equation*}}
\newcommand{\een}{\end{equation*}}
\def \bes{\begin{eqnarray}}
\def \besn{\begin{eqnarray*}}
\def \eesn{\end{eqnarray*}}
\def \ees{\end{eqnarray}}

\newcommand{\Cb}{{\rm \bf C}}

\def \sl2{SL(2,\Cb)}
\begin{document}
\title{Encoding simplicial quantum geometry in group field theories}
\author{{\bf D. Oriti $^a$}\footnote{daniele.oriti@aei.mpg.de} and {\bf T. Tlas $^b$}\footnote{tamer.tlas@aub.edu.lb}}
\affiliation{\small $^a$ Max Planck Institute for Gravitational Physics (Albert Einstein Institute), \\ Am Muehlenberg 1, D-14476 Golm, Germany, EU \\  $^b$ Department of Mathematics, American Univeristy of Beirut, \\ Bliss Street, Beirut, PO Box 11-0236, Lebanon}
\begin{abstract}
We show that a new symmetry requirement on the GFT field, in the context of an extended GFT formalism, involving both Lie algebra and group elements, leads, in 3d, to Feynman amplitudes with a simplicial path integral form based on the Regge action, to a proper relation between the discrete connection and the triad vectors appearing in it, and to a much more satisfactory and transparent encoding of simplicial geometry already at the level of the GFT action. \end{abstract}

\maketitle
\noindent
\section{Introduction}
\ni The field of non-perturbative quantum gravity is progressing fast
\cite{libro}, in several directions. Group field theories \cite{iogft,iogft2, laurentgft} are one of them, and represent a point of convergence of several other formalisms and ideas for defining a complete theory of spacetime dynamics at the microscopic level. 
Group field theories are quantum field theories over group manifolds characterized by a non-local pairing of field
arguments in the action, and can be seen as a generalization of
matrix models \cite{mm} that have proven so useful for our
understanding of 2d quantum gravity (and string theory) (and of
the subsequent, but less developed, tensor models
\cite{gross}). The combinatorics of the field arguments in
the interaction term of the GFT action matches the combinatorics
of (d-2) faces of a d-simplex, with the GFT field itself
interpreted as a (second) quantization of a (d-1)-simplex. The
kinetic term of the action, in turn, dictates the rules for gluing
two such d-simplices across a common (d-1)-simplex, and thus for
the propagation of (pre-)geometric degrees of freedom from one to
the next. See \cite{iogft,iogft2, laurentgft} for details. Because of this
combinatorial structure, the Feynman diagrams of a GFT are dual to
d-dimensional simplicial complexes. The field arguments assigned to these 2-complexes the
same group-theoretic data that characterize spin foam models, and,
most importantly, the GFT perturbative expansion in Feynman
amplitudes define uniquely and completely a spin
foam model.
Spin foam models \cite{SF}, in turn, provide a covariant formulation
of the dynamics of loop quantum gravity \cite{LQG}, adapted to a simplicial context, and a new
algebraic version of the discrete quantum gravity approach based
on simplicial path integrals, as for example Regge calculus \cite{williams}
and dynamical triangulations \cite{DT}. In fact, the spin foam amplitudes of the most studied models can be related more or less directly to simplicial gravity path integrals, weighted by the Regge action.  It can then be argued
\cite{iogft,iogft2,gftfluid} that GFTs represent a common
framework for both the loop quantum gravity/spin foam approach and
simplicial approaches, like quantum Regge calculus and (causal)
dynamical triangulations. On top of this, they offer new tools for addressing the outstanding issue of the continuum approximation of the discrete (pre-)geometric quantum structures used in such approaches \cite{gftfluid}.

\subsection*{The extended GFTs}
\ni The link between the Feynman amplitudes of several group field theories, i.e. several spin foam models, for quantum gravity in 3 and 4 dimensions with simplicial gravity is known. In particular, in 3 dimensions we know that the Ponzano-Regge spin foam model, which is obtained as the Feynman amplitude of the Boulatov group field theory, can also be derived from and is equivalent to the partition function of topological BF theory discretized on the simplicial complex dual to the Feynman diagram of the Boulatov model. Moreover, and this can now be understood as a consequence of the first fact, it is known that the  and quite well understood in the large spin limit the $6j$-symbol, which is the basic building block of the same Ponzano-Regge model, is equal to the cosine of the Regge action for simplicial gravity with the spins giving the edge lengths and thus the metric of the simplicial complex. In 4 dimensions, most of the known spin foam models are constructed based on the formulation of gravity as a constrained BF theory, but a complete derivation of them from a simplicial path integral is lacking. On the other hand,  the large spin limit of the vertex amplitudes of these models can again be related to simplicial gravity actions, suggesting once more a simplicial path integral representation of the full GFT Feynman amplitudes. \\

\ni However, the precise way in which group field theories  encode simplicial geometry in their action and Feynman amplitudes is unclear. This is due to the fact that we lack a representation or a formulation of GFTs based on or involving explicitly discrete metric variables, being the edge lengths, or the Lie algebra elements representing the B  field of BF theory (the gravity triad in 3d, the wedge product of gravity tetrads in 4d) at the simplicial level. Missing this representation, the geometric information coded in the kinetic and vertex terms of the GFT action is rather obscure, the only clear aspects of it being those that can be read out easily in the connection (group) variables or in representation variables, corresponding to the two existing representations of GFTs. 
We would also expect that an appropriate formulation of GFTs based on metric variables, e.g. the discrete B variables of simplicial BF, would give Feynman amplitudes with the explicit form of simplicial path integrals for some gravity action, that can then be turned, if necessary or useful, into spin foam form by expansion in group representations. \\

\ni A recent line of development in group field theory research \cite{feynman, generalised, newgfts} can be understood as the attempt to define such a formulation involving metric variables and producing Feynman amplitudes with the explicit form of simplicial gravity path integrals. In particular, in \cite{newgfts} (to which we refer for more details) an extended GFT formalism has been proposed, characterized by two main features: 1) the field is defined as a function of two sets of variables: group elements representing the parallel transport of the gravity (or BF) connection and Lie algebra elements representing the discrete $B$ variables of Bf theory; this doubling of variables is meant to facilitate the encoding of their respective and relative geometric relation at the level of the action; 2) a non-trivial kinetic term  given by the difference (up to constants) between the Laplacian-Beltrami differential operator acting on the group manifold and the modulus square of the $B$ variables, one for each of the $D$ arguments of the field, in $D$ dimensions, thus one for each of the $(D-2)$-simplices in the simplicial complex dual to each Feynman diagram. The specific form of the resulting Feynman amplitudes is mainly due to this choice of kinetic term. Its main feature is that it relaxes the identification between the $B$ variables and the left derivatives acting on the group manifold, and thus the conjugate nature of $B$ variables and connection group elements. This implies a departure from BF theory, but it is necessary in order to maintain a simultaneous dependence of the two sets of variables.\\

\ni The main nice feature of the new formalism is that indeed the Feynman amplitudes take the form of a simplicial path integral for a gravity action, and that several features of the simplicial geometry seem to be correctly implemented in the GFT action and in the same amplitudes. However,  both the form of the amplitudes and some aspects of the relation between connection and metric variables are not satisfactory, as we are going to discuss in the following, suggesting that the simplicial geometry is still not correctly implemented in the GFT action used in \cite{newgfts}. In this paper, we improve on the situation, and show that a different symmetry requirement on the GFT field, relating more strictly the two sets of variables, leads to a simplicial gravity path integral which is much closer to the desired one and to a better encoding of simplicial geometry in the GFT action. We focus on the 3d case. The 4d case is dealt with in \cite{new4d}.

\section{Encoding simplicial geometry in GFT: the 3d model}
\subsection{A modified symmetry requirement for the GFT field}
\ni While retaining the same choice of dynamics (action) as in the model introduced in \cite{newgfts}, which we will recall below, we introduce a different symmetry requirement on the GFT field. We stipulate that the GFT field should satisfy

\be
\label{eq:invariance}
\phi ( g_1 \, h^{-1} , g_2 \, h^{-1} , g_3 \, h^{-1} ; h B_1 h^{-1}, h B_2 h^{-1}, h B_3 h^{-1}) = \phi (g_1 , g_2, g_3 ; B_1, B_2, B_3).
\ee

\begin{figure}[htbp]
\centering
\includegraphics[width=0.35\textwidth]{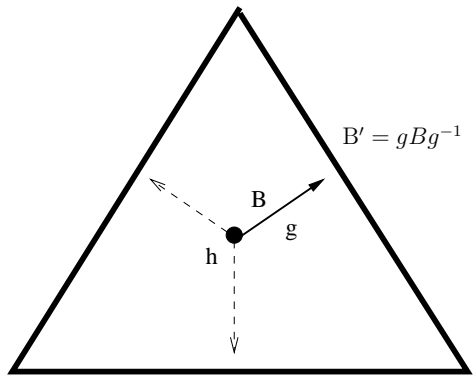}
\caption{\textit{This is a graphical representation of the GFT field. The group element $g$ is the parallel transport going from the centre of the triangle to the edge (as is indicated by the arrow). The Lie algebra element $B'$ is the $B$ field discretized on the edge as seen in the frame of the edge, while the element $B$ is the Lie algebra element as `seen' from the centre of the triangle. Finally, $h$ is the group element encoding the freedom of choosing a frame at the centre of the triangle.}}
\label{fig:field}
\end{figure}

\ni The motivation for the new symmetry requirement is the geometric interpretation of the field itself and of its arguments.  In GFT the field is interpreted as a 2nd quantized triangle, according to the general interpretation and construction of GFT models \cite{iogft}. Then the three $g$'s represent the three parallel transports of a gravity connection, going from the centre of the triangle to the centres of its three edges as is depcited in figure \ref{fig:field}. The interpretation of the $B$'s is slightly more subtle : they represent the parallel transports of the continuum $B$ field, which is a (triad) 1-form, averaged over the edges of the triangle $t$ to which the field $\phi$ corresponds, to the centre of the triangle using the $g$'s, and thus now expressed in the reference frame associated to the same triangle. In other words, if $B_1 = \int_{\txt{\footnotesize{edge 1}}} B(x)$, then $B_1(t) = g_1^{-1} B_1 g_1$\footnote{The reason why the group elements act on the $B$'s via conjugation follows from the fact that the $B$ field in BF theory is a section of the bundle associated to the principal one via the adjoint representation.}.\\

\ni Then the new symmetry imposes the invariance of the model under local rotations of the reference frame in each triangle $t$, which are effected by the group element $h$. Indeed, in view of the interepretation given above for the $B$ variables, when $g \rightarrow g h^{-1}$ (which is exactly how a parallel transport changes when one performs a rotation at the starting point) then $B(t) \rightarrow h B(t) h^{-1}$. We will see in the next section, moreover, that we can identify the new reference frame in which the $B(t)$ variables are expressed, after such rotation, as the (arbitrary) reference frame located at the center of each tetrahedron, corresponding to the interaction vertex of the GFT action, in the Feynman expansion of the model \cite{iogft}, and interpret accordingly the group elements $h$ as the parallel transports of the gravity connection from the center of the triangle to the center of the tetrahedron. \\

\ni This interpretation of the $B$ variables (and of the group elements $g$ and $h$) is confirmed by the expression of the Feynman amplitudes of our model, which take the form of simplicial path integrals, and it is also supported by the canonical analysis of the discrete BF theory performed in \cite{EPR, biancajimmy}. In the last of these works, the fact that the $B$ variables involve a connection in their very definition shows up in their noncommutativity with respect to the BF Poisson brackets; a similar non-commutativity is present also in continuum canonical Loop Quantum Gravity \cite{ACZ}, after smearing of the basic triad field variables using smearing functions with support on 1-dimensional submanifolds of the canonical slice, and such smearing can be understood as analogous to the discretization procedure of BF theory and gravity underlying the GFT approach and our model\footnote{This noncommutativity of the discretized $B$'s can also be understood to be a consequence of the Jacobi identity \cite{biancajimmy, ACZ}.}.\\

\ni In order to show one immediate consequence of our modified symmetry requirement, let us harmonically expand the field, which is a function of the group elements, in representations $J_i$ of $SU(2)$ (and corresponding vector indices $\alpha_i,\beta_i$, using the Peter-Weyl decomposition:

\bes
\label{eqnarray:projection}
\phi( g_i ; B_i ) & = & \int dh \tilde{\phi}( g_i h^{-1} ; h B_i h^{-1}),  \\
\nonumber
& = & \int dh \,\, \bigg ( \prod_{i=1}^3 \sum_{J_i, \alpha_i, \beta_i, \gamma_i} \bigg ) \, \tilde{\phi}^{J_i}_{\alpha_i \gamma_i}( h B_i h^{-1}) \prod_{i=1}^3 D^{J_i}_{\alpha_i \beta_i}(g_i) D^{J_i}_{\beta_i \gamma_i}( h^{-1}),\\
\label{eqnarray:harmoniccov}
& = & \bigg ( \prod_{i=1}^3 \sum_{J_i, \alpha_i, \beta_i} \bigg ) \phi^{J_i}_{\alpha_i \beta_i}(B_i) \prod_{i=1}^3 D^{J_i}_{\alpha_i \beta_i}(g_i) ,
\ees

where, in the first equation above we've implemented the symmetry of the field via a projection of a generic non-symmetric field $\tilde{\phi}$, and have defined $\phi^{J_i}_{\alpha_i \beta_i}(B_i)$ to be

\ben
\phi^{J_i}_{\alpha_i \beta_i} (B_i) = \sum_{\gamma_i} \int dh \tilde{\phi}^{J_i}_{\alpha_i \gamma_i}(h B_i h^{-1}) \prod_{i=1}^3  D^{J_i}_{\beta_i \gamma_i}( h^{-1}).
\een

\ni Now, note that, if $\tilde{\phi}^{J_i}_{\alpha_i, \gamma_i}( h B_i h^{-1})$ did not depend on $h$, $\phi^{J_i}_{\alpha_i \beta_i} (B_i)$ would be an intertwiner, i.e. an invariant tensor, in both of its sets of indices. Now, on the other hand it is easy to see that 

\ben
\sum_{\alpha_i} \bigg ( \prod_{i=1}^3 D^{J_i}_{\delta_i \alpha_i}(k) \bigg ) \phi^{J_i}_{\alpha_i \beta_i} (B_i) = \phi^{J_i}_{\alpha_i \beta_i} (k B_i k^{-1}).
\een

\ni In other words, $\phi^{J_i}_{\alpha_i \beta_i} (B_i)$ is not an invariant tensor (intertwiner) but rather a covariant one. As a consequence, the spinfoam and the boundary states obtained from this theory will not be invariant under local rotations acting on the connection by itself, but rather covariant (invariant under local rotations acting simultaneously on the connection and the $B$ field). A similar covariance property has been advocated by Alexandrov starting from the canonical analysis of the theory (e.g. \cite{alexandrov}). 

\subsection{GFT action and geometric meaning of the vertex function}
\ni The action for the model is:

\bes
\nonumber
& S [ \phi] & =  \frac{1}{2 (2 \pi)^9} \int_{SU(2)^3 \times \mathfrak{su}^3(2)} \bigg ( \prod_{i=1}^3 \, dg_i \, dB_i \bigg )  \, \phi( g_i ; B_i) \, \Big (  B_i^2 + \Box_{SU(2)} - \frac{1}{8} \Big ) \, \phi( g_i ; B_i) + {}  \\
& & + \frac{\lambda}{24 (2 \pi)^{36}} \int_{SU(2)^{12} \times \mathfrak{su}(2)^{12}} \bigg ( \prod_{i \neq j =1}^{12} \, dg_{ij} \, dB_{ij} \bigg ) \Bigg [ \prod_{i < j} \delta ( g_{ij} g_{ji}^{-1} ) \, \delta ( B_{ij} - B_{ji} ) \Bigg ] \, \phi (g_{1j} ; B_{1j}) \dots \phi (g_{4j} ; B_{4j}).
\ees

\ni The first summand is the kinetic (quadratic) term and the second one is the interaction. They are given by exactly the same formulae as in \cite{newgfts} (having set the arbitrary constant $m^2$ in the kinetic term, appearing in the general formulation of \cite{newgfts}, to the identity). The only difference between the model put forward here and the one in \cite{newgfts} is in the new symmetry property of the field $\phi$.\\

\ni One can gain more insight into the geometric meaning of new symmetry of the field by rewriting the vertex term in terms of the un-projected fields (\ref{eqnarray:projection}) and changing variables

\ben
\txt{Vertex} \sim \int_{SU(2)^{12} \times \mathfrak{su}(2)^{12}} \bigg ( \prod_{i \neq j =1}^{12} \, dg_{ij} \, dB_{ij} \, dh_i \bigg ) \Bigg [ \prod_{i < j} \delta ( g_{ij} h_i h_j^{-1} g_{ji}^{-1} ) \, \delta ( h_i^{-1} B_{ij} h_i - \, h_j^{-1} B_{ji} h_j ) \Bigg ] \, \tilde{\phi} (g_{1j} ; B_{1j}) \dots \tilde{\phi} (g_{4j} ; B_{4j}).
\een

\ni Consider now the expression in the square brackets

\be
\label{eq:newvertex}
\prod_{i < j} \delta ( g_{ij} h_i h_j^{-1} g_{ji}^{-1} ) \, \delta ( h_i^{-1} B_{ij} h_i - \, h_j^{-1} B_{ji} h_j ) .
\ee

\ni To understand this object, recall the geometric interpretation of the variables we are dealing with, mentioned above (and in \cite{new4d}). Thus, $g_{ij}$ represents the parallel tranport from the centre of the i-th triangle to its edge common with the j-th triangle. $B_{ij}$ is the $B$ field averaged on the edge common between the i-th and j-th triangles as seen in the reference frame located at the centre of the i-th triangle. $h_i$ represents the parallel transport from the centre of the 3-simplex dual to the interaction vertex (formed by the four triangles represented by the four fields in the same interaction) to the centre of the i-th triangle.\\
With this interpretation in mind, the formula (\ref{eq:newvertex}) acquires a simple geometric meaning. It is simply a product of six pairs of delta functions, one pair per wedge of the simplex, enforcing two geometric conditions:

\begin{itemize}
\item the flatness of the wedge (portion of a dual face located inside the tetrahedron) (this is the first delta in (\ref{eq:newvertex}). This requirement (that the parallel transport around the wedge is trivial) is standard and present in all the GFTs put forward to date; it encodes in the GFT action the piecewise flat geometry characterizing the corresponding Feynman amplitudes, with generic discrete geometries obtained by gluing together flat simplices, as in (quantum) Regge calculus \cite{williams}.
\item the compatibility of the connection with the triad (the $B$ field). This is the second delta in (\ref{eq:newvertex}). It enforces that the $B$ variables corresponding to the same edge as seen in the (reference frames of the) two triangles sharing it coincide up to parallel transport, and that this parallel transport is performed by means of the discrete connection $h$. In particular, they coincide when transported to the centre of the 3-simplex, by means of the same connection. 
\end{itemize}

\begin{figure}[htbp]
\centering
\includegraphics[width=0.5\textwidth]{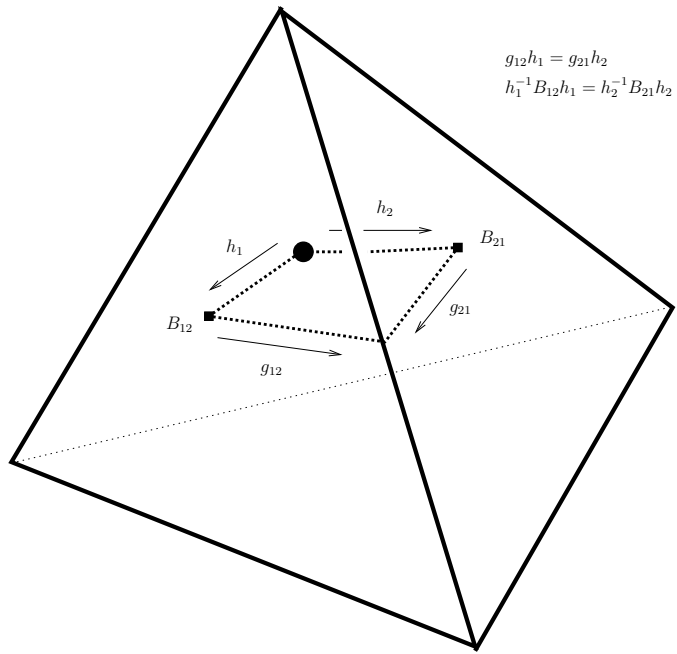}
\caption{\textit{This is the graphical representation of the new vertex. We have depicted only one wedge of this vertex. The arrow indicate the orientation of the parallel transports. The $B$'s are best thought of as being located at the small black squares. The two relations at the top right corner are those enforced by the new vertex.}}
\label{fig:vertex}
\end{figure}

\ni See the figure Fig. \ref{fig:vertex} for a graphical representation of the vertex function.\\

\ni Let us remark that the second condition (the compatibility of the $B$ field with the connection) has been recently obtained directly through the canonical analysis of the discrete theory \cite{biancajimmy}, and its meaning and importance have been stressed in \cite{valentin}, which lends even more support to imposing the new symmetry on the GFT field. Conversely, this simple GFT implementation shows that this requirement is the natural counterpart of the flatness of each 3-simplex, as measured by the connection variables of the GFT field, if one recalls that both the connection and the $B$ variables transform simultaneously under the rotation gauge group of the theory. \\

\ni Before we proceed with the quantization of the model above, let us note that due to the wedge flatness one could have replaced the $h$'s in the second delta with the $g$'s. In other words, instead of (\ref{eq:newvertex}) one could write

\ben
\prod_{i < j} \delta ( g_{ij} h_i h_j^{-1} g_{ji}^{-1} ) \, \delta ( g_{ij} B_{ij} g_{ij}^{-1} - \, g_{ji} B_{ji} g_{ji}^{-1} ) ,
\een

i.e. instead of stipulating that the $B$ in one triangle is equal to the $B$ in the other up to a parallel transport passing through the centre of the simplex, we can, because of the flatness of the wedge, stipulate that the $B$ in one triangle is equal to the $B$ in the other up to a parallel transport going through the centre of the edge of the simplex. This encodes the fact that the two $B$ variables in each field (triangle) come from the {\it same} Lie algebra element located at the common edge, by means of parallel transport. The two expressions lead to identical Feynman amplitudes. 

\subsection{Feynman amplitudes}
\ni We define the partition function of the model using its perturbative expansion in Feynman diagrams:
\be
Z\,=\,\sum_\Gamma\frac{\lambda^{N_\Gamma}}{sym(\Gamma)}\,Z_\Gamma\;\;\;\; .
\ee 
where $Z_\Gamma$ are the Feynman amplitudes associated to the diagram $\Gamma$, with number of interaction vertices $N_\Gamma$ and order of automorphism group $sym(\Gamma)$.
These are given \cite{iogft}, by construction, by cellular complexes (or fat graphs) dual to 3-dimensional simplicial complexes. 
Let us now calculate the corresponding Feynman amplitudes. To do this we need the two building blocks which make the amplitudes, the propagator and the vertex amplitude. We have already isolated and discussed the vertex function (\ref{eq:newvertex}).\\

\ni The propagator is given by the following expression

\ben
D_F [g_i, \tilde{g}_i; B_i, \tilde{B}_i  ] = \int_{SU(2)} dh \prod_{i=1}^3 \bigg ( K \Big[ g_i h \tilde{g}_i^{-1} , B_i^2 - \frac{1}{8}  \Big ] \delta( \tilde{B}_i - h B_i h^{-1} ) \bigg ),
\een

where $K[.,.]$ is just the Feynman propagator for (variable) mass $B_i^2 -\frac{1}{8}$, which in turn can be computed using the Schwinger proper time representation in terms of the Schroedinger kernel) on the group manifold $SU(2)$ \cite{newgfts,camporesi}. The only difference between this expression and the propagator used in \cite{newgfts} is that the $B$'s are equated up to parallel transport (the presence of the $h$ in the delta function relating the $B$'s). Obviously, this parallel transport does not affect the \lq\lq mass\rq\rq $B^2$.\\

\ni With the two needed ingredients at hand one can proceed and calculate the Feynman amplitudes of the quantum theory whose classical action is given above. As is always the case one obtains a fat graph which can be understood as the 2-complex locally dual to a three dimensional simplicial complex. The amplitudes can be factorized per dual face (or per edge of the triangulation) and are given by

\ben
Z_\Gamma = \int_{SU(2)^E} \bigg(  \prod_{\txt{dual edges}} dh_e \bigg ) \int_{\mathfrak{su}(2)^F} \bigg ( \prod_{\txt{dual faces}} dB_f \bigg ) \, \prod_{\txt{dual faces}}\, A_{N_f} [ H_f , B_f] ,
\een

where $E$ is the number of dual edges. $F$ is the number of dual faces. $N_f$ is the number of vertices in the dual face $f$, $h_e$ is the group element associated to the dual edge $e$, $B_f$ is the Lie algebra element associated to the dual face $f$. $H_f$ is the holonomy around the dual face $f$ (i.e. obtained by taking the product of the corresponding $h$'s). Note that the various  $B$ variables associated to the same edge of the triangulation as seen by the various triangles sharing it can be integrated out, leaving a dependence on only a single $B_f$, and on the associated holonomy, i.e. the $B$ variable in the frame of the triangle used as starting point for the definition for the same holonomy (one can easily check that the amplitudes are then independent on the choice of this starting point). Finally, $A_{N_f} [ H_f , B_f]$ is given by:

\be
\label{eq:amplitude}
A_{N_f} [ H_f , B_f] = \frac{\delta(B_f - H_f B_f H_f^{-1})}{(N_f - 1)!} \int_{\mathbb{R}} dT \, e^{i (B_f^2 - \frac{1}{8}) T} \, \theta(T) T^{N_f -1} K [ H_f , T ]  ,
\ee

where $K[ H_f , T]$ is the Schroedinger kernel for the group manifold $SU(2)$ in proper time \cite{marinov}:

\ben
K [ H_f , T] = \frac{1}{(4 \pi i T)^{\frac{3}{2}}} \sum_{n=-\infty}^{\infty} \bigg ( \frac{\theta_f + 4 \pi n}{2 \txt{sin}(\frac{\theta_f}{2})} \txt{Exp} \Big [ \frac{i}{2T} \big( \theta_f + 4 \pi n \big)^2 + \frac{iT}{8} \Big ] \bigg )   .
\een

\ni Notice the explicit dependence on the combinatorial structure of the underlying simplicial complex, though the number $N_f$ of dual vertices in each dual face . No such explicit dependence is present in the usual BF path integral, known to be (after regularization) triangulation independent, for given simplicial topology. 

\ni The integral in (\ref{eq:amplitude}) can now be done explicitly and the result is

\be
\label{eq:amplitude2}
A_{N_f} [ H_f , B_f ] = \mu ( [ \theta_f ] , | B_f | , N_f )  \, \delta(B_f - H_f B_f H_f^{-1})\,  e^{ i | B_f| | [  \theta_f]|} ,
\ee

where we have dropped, for transparency, the explicit sum enforcing the periodicity of the expression, which is now signified by the square brackets around the $\theta_f$.\footnote{For details on the notation, we refer to \cite{newgfts}.} This sum, whose meaning is discussed in the next subsection, is always understood to be present at the leftmost position of any expression containing $\theta_f$. The $\mu ( [ \theta_f ] , | B_f | , N_f )$ in the above expression is given by

\ben
\mu ( [ \theta(H_f) ] , | B_f | , N_f ) = \frac{- i \sqrt{2}}{16 \pi (N_f -1)!} \frac{1}{\txt{sin}(\theta_f)} \bigg (\frac{ [ \theta_f ]}{|B_f|} \bigg)^{N_f -1} \sum_{K=0}^{N-2} (-1)^K \frac{(N+K-2)!}{K! (N-K-2)!} \frac{1}{(2 i |B_f| |[\theta_f]|)^K}.
\een

\ni The amplitude for the whole two complex is then given by

\be
\label{eq:partition}
Z_\Gamma = \int_{SU(2)^E} \bigg(  \prod_{\txt{dual edges}} dh_e \bigg ) \int_{\mathfrak{su}(2)^F}  \prod_{\txt{dual faces}} \bigg ( dB_f \, \, \delta(B_f - H_f B_f H_f^{-1})  \, \, \mu( B_f , H_f) \bigg ) \,\, e^{i S},
\ee

\ni The measure term $\mu$ can be re-written in terms of its modulus and phase, with the phase interpreted as defining quantum correction to the action term
$S=\sum_f | B_f| | [  \theta_f]|$. We will discuss this action term in the following, while the measure and the corresponding quantum corrections to the action are discussed in \cite{newgfts}. Note that (\ref{eq:amplitude2}) is exactly the same amplitude as for the 3d model in \cite{newgfts}, apart from the new delta function,  which is the only consequence of the modified symmetry requirement satisfied by the field. 

\subsection{Properties of the model}
\ni Let us now study the properties of these Feynman amplitudes. In particular, as anticipated, we focus on the action $S$ and on the new delta function appearing in each dual face contribution to the amplitude.

\ni The action appearing in the amplitude is:
\be
\label{eq:discrete action}
S = \sum_{\txt{dual faces}} |B_f| | [ \theta_f  ] |,
\ee

in which the $|B_f|$ can be interpreted as the length of the edge dual to the dual face $f$ and $\theta_f$ is the angle of curvature corresponding to the holonomy $H_f$ along the boundary of the dual faces around the same edge. Let us now try to get a clearer understanding of this action, and to compare it with the two most common discretizations of BF theory on a simplicial lattice. In order to do so, we take a step back and consider the two main ways the BF action is usually discretized.\footnote{Recently, there has been a third discrete action proposed in \cite{laurentflorian} (see also \cite{eterajimmy}). This new one, as far as ensuing discussion is concerned, can be considered to be a variation of (\ref{eq:actiongroup}).} The first is given by

\be
\label{eq:actiongroup}
S_{BF}^{1} = \sum_{\txt{dual faces}} \txt{Tr}(B_f H_f),
\ee

where $B_e$ is the Lie algebra element associated to the dual face $f$ and $H_f$ is the holonomy around the same dual face. The trace is taken in the fundamental representation.\\

\ni Another possible discretization \cite{kirilllaurent} is given by

\be
\label{eq:bf action}
S_{BF}^2 = \sum_{\txt{dual faces}} \txt{Tr}(B_f \txt{ln}(H_f)),
\ee

where $F_f = \txt{ln}( H_f) = [\theta_f(H_f)] \vec{n_f}(H_f) $ denotes the inverse of the exponential map applied to the holonomy $H_f$ and identifies the corresponding curvature in terms of a total rotation angle $\theta_f$ and a direction of rotation $\vec{n_f}$. Now, as is the case in the complex plane, the logarithm is a multivalued function here as a consequence of the compactness of SU(2). Therefore one is faced with the question of which branch of the logarithm to choose, and since there is no apriori reason to prefer one over the other, the only way out is to sum over all of them with equal weights. Equivalently, one can say that, in defining the action above, one has to compute the geodesic distance, corresponding to the angle of rotation of the group element $H_f$, on the group manifold $SU(2)$; being this a compact manifold, there is a 1-parameter family of geodesics that can be used to compute such distance, parametrised by an integer $n$, ad one has to sum over this parameter to treat all such geodesics on equal footing. This is exactly what is achieved by the periodicity sum in (\ref{eq:partition}), and that is represented by the notation $[\theta_f]$ denoting the {\it equivalence class} of rotation angles. An alternative way of looking at the situation is to take the fact that the log is a multivalued function seriously and thus its natural domain is a covering space of SU(2) and not SU(2) itself, just like the natural domain of the complex log is a non-trivial Riemann surface and not just a plane (with a cut).\footnote{One might wonder whether this problem can be avoided by using Lie algebra elements from the start instead of group elements, as arguments of the GFT field. We do not pursue further this possibility here.} Summarizing, the periodicity sum in (\ref{eq:partition}) is a consequence of the multivaluedness of the logarithm which appears (in disguise - this is the angle $\theta$) in the expression for the heat kernel and thus is naturally integrated not on the group but on the corresponding covering space \cite{camporesi}. This issue has been discussed also in \cite{newgfts}. In any case, we see that our action (\ref{eq:discrete action}), for the variables that appear in it, and because of the need to consider the equivalence class of rotation angles, corresponds to a discretization of BF theory of the type of (\ref{eq:bf action}). \\

\ni But the two actions still look different. Let us rewrite (\ref{eq:bf action}) denoting $\txt{ln}(H_f)$ by $F_f$ and using the isomorphism between $\mathfrak{su}$(2) and $\mathbb{R}^3$ as vector spaces (so that the trace is the standard scalar product on $\mathbb{R}^3$): 

\besn
S^2_{BF} & = & \sum_{\txt{dual faces}} \vec{B}_f \cdot \vec{F}_f.
\eesn

\ni Now, while (\ref{eq:bf action}) involves the scalar product of the two vectors $\vec{B}_f$ and $\vec{F}_f$, our action (\ref{eq:discrete action}) depends only on the magnitudes of these same two vectors ($\theta_f = |\vec{F}_f|$), with the information on their relative direction missing. Moreover, as far as the two actions are concerned, the two theories have different symmetries with respect to local rotations. The BF theory action is invariant under local rotations acting simultaneously on both the $B$ field and the connection, while the theory given by the action (\ref{eq:discrete action}) is invariant under a larger symmetry group given by {\it independent} local rotation of the $B$ field and the connection. 

\ni However, the situation is drastically modified by the new delta function appearing in each dual-face amplitude (\ref{eq:amplitude2}). This part of the amplitude is \textit{not} invariant under independent rotations of the two variables but only under simultaneous ones (i.e. when we transform the $B_f$ and the $H_f$ by conjugation), thus restoring the correct symmetry to the theory. Moreover, this same delta function can be understood as  fixing the relative direction of curvature and the triad B vector, and thus leading from the BF action (\ref{eq:bf action}) too our action (\ref{eq:discrete action}). \\

\ni Let us show how this happens by analyzing this delta function. To simplify the formulas below we will denote the unit vector along $\vec{F}_f$ by $\vec{n}_f$, i.e. $\vec{n}_f = \frac{\vec{F}_f}{|\vec{F}_f|}$. Then

\besn
\delta(B_f - H_f B_f H_f^{-1}) & = & \delta \Big [ \vec{B}_f -  \big (\vec{B}_f  + \txt{sin}(2 \, |\vec{F}_f| ) \, \vec{n}_f \times \vec{B}_f      + (1 - \txt{cos} (2 \,|\vec{F}_f|)) \, \vec{n}_f \times ( \vec{n}_f \times \vec{B}_f)  \big )       \Big ]= \\
& = & \frac{1}{4 \pi (\txt{sin}^2(2 \, |\vec{F}_f| ) \, ( \vec{n}_f \times \vec{B}_f)^2      + (1 - \txt{cos} (2 \,|\vec{F}_f|))^2 \, (\vec{n}_f \times ( \vec{n}_f \times \vec{B}_f))^2)}   \times {} \\ 
&& {} \qquad \qquad \times \delta \Big [ \txt{sin}^2(2 \, |\vec{F}_f| ) \, ( \vec{n}_f \times \vec{B}_f)^2      + (1 - \txt{cos} (2 \,|\vec{F}_f|))^2 \, (\vec{n}_f \times ( \vec{n}_f \times \vec{B}_f))^2 \Big ]=\\ \\
& = & \frac{1}{4 \pi ( (2  - 2 \txt{cos}(2 |\vec{F}_f|) )\, ( \vec{n}_f \times \vec{B}_f) ^2)}   \delta \Big [  \sqrt{  (2  - 2 \txt{cos}(2 |\vec{F}_f|) )\, ( \vec{n}_f \times \vec{B}_f) ^2   } \Big ]=\\ \\
& = & \frac{1}{4 \pi \,  4 \, \txt{sin}^2(|\vec{F}_f|) \, (\vec{n}_f \times \vec{B}_f)^2 }  \delta \Big [ 2 \, \txt{sin}(|\vec{F}_f|) \, |(\vec{n}_f \times \vec{B}_f)  |  \Big ]= \delta \Big [  2 \, \txt{sin}(|\vec{F}_f|) \, \vec{n}_f \times \vec{B}_f \Big ].
\eesn

\ni From the last formula it is clear that the delta function is non-zero only when either $B_f$ or $F_f$ are zero (corresponding to a degenerate and to a flat geometry, respectively), or when they are aligned with each other, in which case their vector product $\vec{B}_f \times \vec{F}_f$ is zero. This is also clear geometrically as a non-zero vector ($B_f$) will be preserved by a non-trivial rotation ($H_f$) only when the axis of rotation $F_f$ is aligned with the vector. Notice that the delta function can be turned from a delta function on the algebra $\mathfrak{su}(2)$, thus constraining the Lie algebra elements $B_f$, to a delta function on the group $SU(2)$, and thus constraining the vonnection group elements $h_e$. Excluding degenerate configurations, the general solution of the constraint imposed on $(B,h)$ configurations by the delta function, seen as a constraint on the group elements $h_e$, is then: $H_f = e^{i \theta_f \frac{B_f}{|B_f|}}$. Notice that this restriction on the curvature holonomies follows simply by the condition that the same group elements are used to parallel transport the triad variables $B_f$ from a reference frame to another across the simplicial complex, a condition encoded in our GFT action by the symmetry requirement under rotations on the field. The same condition on bivectors and/or holonomies, following a similar rule for parallel transport of B variables, has been identified and studied in \cite{valentin}, in both 3 and 4 dimensions. The effect of this symmetry requirement and the constraints on bivectors and holonomies resulting from it in 4d have been discussed instead in \cite{new4d}, in the context of the 4d counterpart of the GFT model presented here. From this re-writing it is also clear why the delta function is only invariant under simultaneous rotations of $\vec{B}_f$ and $\vec{F}_f$.\\

\ni Now, keeping in mind that in the partition function the integration is restricted to those configurations for  which every $B_f$ is aligned with the corresponding $F_f$ it should be clear that there is almost no difference between $\vec{B}_f \cdot \vec{F}_f$ and $|\vec{B}_f| \, |\vec{F}_f|$. The remaining difference is that the sign of the first expression, after imposition of the alignment  condition, can be positive and negative while that of the other one can only be positive. Put differently, given that the integration range is restricted to configurations where there is alignment between the $B$'s and the $F$'s we see that the GFT gives amplitudes which are equivalent to those of the BF theory with an extra condition, that $\vec{B}_f \cdot \vec{F}_f \geq 0$. This is of course nothing but the causality condition which was proposed in \cite{causalmatter3d}. \\ 

\ni Thus we conclude that (\ref{eq:partition}) can be equivalently written as

\ben
Z = \int_{SU(2)^E} \bigg(  \prod_{\txt{dual edges}} dh_e \bigg ) \int_{\mathfrak{su}(2)^F}  \prod_{\txt{dual faces}} dB_f \prod_{\txt{dual faces}} \bigg ( \Theta(B_f \cdot F_f) \, \, \delta(B_f - H_f B_f H_f^{-1})  \, \, \mu( B_f , H_f) \bigg ) \,\, e^{i S_{BF}},
\een

where $\Theta(.)$ is the Heaviside step function. The `alignment' condition (enforced by the delta function) results also in a \lq\lq gauge-fixing\rq\rq for a trivial symmetry  of the BF action, corresponding to shifts of the $B_f$'s in the directions orthogonal to the corresponding $F$'s, i.e. by a Lie algebra element $\phi_f$ (one independent shift per edge of the simplicial complex) satisfying $\phi_f \cdot  F_f = 0$. This transformation is trivialized on-shell, i.e. after the equations of motion (flatness of the holonomy $H_f$) are imposed. As such, it does not correspond to a true gauge symmetry of the theory (hence the quotation marks above), as it does not act at all on solutions, but it is a symmetry transformation of the action nevertheless. Another related interpretation of the above condition on holonomies is as being the discrete analogue of the (part of the) metricity condition on the connection, which represent the second equation of motion of BF theory, $d_\omega B = 0$. This condition, in fact, forces the generic connection $\omega$ to be the torsion-free Levi-Civita connection, a function of the triad (turning the formalism to a 2nd order one, if imposed at the level of the action). At the discrete level, the Levi-Civita connection can be identified with a rotation element lying on the plane orthogonal to the edge of the simplicial complex, and defining the parallel transport of reference frames. We see that our discrete connection satisfies both these requirements. When they are satisfied, the holonomy angle $\theta_f$ can be identified with the deficit angle measuring the curvature of the geometry defined by the discrete triad and the discrete action (\ref{eq:discrete action}) with the Regge action. It remains to be seen as the collection of such conditions for all the holonomies associated to every dual face of the simplicial complex suffices to constrain all the connection degrees of freedom as a function of the triad ones, and thus to turn our initially 1st order theory into a 2nd order one in which the only variables are the lengths of the edges of the simplicial complex, as in ordinary Regge calculus. This is indeed what we would expect to happen if our parallel transport condition relating Lie algebra and group variables is a discretization of the metricity condition for the BF connection.\\

\ni Summarizing, we deduce that modulo the measure factor the proposed GFT generates spinfoam amplitudes for the causal simplicial BF theory, at least in the bulk (i.e. or for a simplicial complex with no boundary), with an additional alignment condition between the connection and the triad, which allows us to interpret the holonomy  rotation angle as a  deficit angle, and the action as a Regge calculus action.\\

\ni Let us calculate the amplitude given by the new GFT for a single simplex to elucidate some more features of the proposed GFT model. This is done by inserting an appropriate observable into the partition function (\ref{eq:partition}). As in any GFT, the observable should be an appropriate gauge invariant functional of the field $\phi$. Here the field is invariant under the diagonal action of $SU(2)$ on both $B$ and $g$ variables, so that, if expressed in terms of a spin network functional (see for example \cite{LQG}), any such observables would imply the use of a \textit{covariant/projected} spin network functional similar to those put forward by Livine \cite{eteraprojected} and developed by Alexandrov and Livine \cite{eteraalex, alexandrov}. Consider the simplest polynomial observable given by a product of fields with arguments paired with the combinatorics of a tetrahedron:

\besn
O_{tet}(\phi)&=&\phi(B_{12},B_{13},B_{14}; g_{12},g_{13},g_{14}) \, \, \phi(B_{12},B_{23},B_{24}; g_{12},g_{23},g_{24}) \times{} \\ & & {}\times \, \,\phi(B_{13},B_{23},B_{34}; g_{12},g_{23},g_{34})\, \, \phi(B_{14},B_{24},B_{34}; g_{14},g_{24},g_{34})\eesn
where we have labeled from 1 to 4 the triangles in the tetrahedron, so that each edge (to which a Lie algebra and a group element are associated) is labelled by a pair of indices $(ij)$ (with $i\neq j$) for the two triangles sharing it, and evaluate its mean value:
\be
\langle O_{tet}(\phi) \rangle_\lambda\, = \, \int \mathcal{D}\phi \, O_{tet}(\phi)\, e^{- S_{GFT}(\phi, \lambda)} 
\ee

to first order in the coupling constant $\lambda$, so that the Feynman diagrams involved in the computation reduce to a single one, obtained by pairing the arguments of the fields in the observable with four propagators and these with a single interaction term, the result being:

\besn
\langle O_{tet}(\phi)\rangle_\lambda & \simeq & \int \prod_i dh_i \, \prod_{(ij)} \Bigg ( \frac{1}{\txt{sin}(\theta_{ij}(h))} \bigg ( \frac{\theta_{ij}(h)}{|B_{ij}|}  \bigg ) 
  \delta \Big ( h_i^{-1} B_{ij} h_i - h_j^{-1} B_{ij} h_j   \Big ) \, \Bigg )  \, \, \,e^{i \sum_{(ij)} | [\theta_{ij}(h)] | |B_{ij}|} \;\;+\;\;O(\lambda^2)\,=\, \nonumber \\  &&\hspace{-1.5cm} = \int \prod_i dh_i \int \prod_{(ij)}dG_{ij}\,\prod_{(ij)} \Bigg (\frac{\theta_{ij}(G)}{|B_{ij}|\txt{sin}(\theta_{ij}(G))} 
  \delta \Big ( B_{ij} - (h_i^{-1} h_j) \triangleright B_{ij}   \Big ) \,  \,\delta(G_{ij}h_i^{-1}h_j)  \Bigg )\, \,e^{i \sum_{(ij)} | [\theta_{ij}(G)] | |B_{ij}|} \;\;+\;\;O(\lambda^2),
\eesn

where $\theta_{ij}(h)$ is the angle of rotation of the group element $h_i h_j^{-1}$, which can be equivalently expressed in terms of the group element $G_{ij}$ by inserting an appropriate delta function equating it to the group element $h_i h_j^{-1}$. \\

\ni Let us discuss briefly this expression.

\ni First, notice that the exponential factor containing the action term arises from the \textit{propagators} of the model, as the vertex amplitudes are just delta functions. Thus, even for a single tetrahedron, it is the four propagators, which contain the non-trivial contribution to the dynamics.\\

\ni The geometric meaning of the various contributions and of the variables appearing in the amplitude is apparent (compare it for example to the discrete action and amplitude for BF theory on a single simplex in \cite{EPR}). The amplitude is a function of the triad variables $B_{ij}$ assigned to the boundary edges of the tetrahedron, while the dependence on the group elements $g$ (arguments of the boundary state) having dropped out of the expression. The underlying reason is that the dynamics depends only on the {\it difference}, between the group elements that are assigned to the boundary edges by the boundary state. The other variables appearing in the amplitude are the group elements $h_i$, once more paying the role of the discrete connection in the bulk, and the group elements $G_{ij}$; they can be interpreted as representing the discrete connection on the boundary of the simplicial complex, which are forced to be a function of the bulk one by the flatness condition on each wedge (imposed by the delta function). Now, the other delta functions constraining the variables associated to each edge come once more from the symmetry we have imposed on the GFT field. They force the connection $h_ih_j^{-1}$ (or equivalently, $G_{ij}$) to lie in the plane orthogonal to the edge $(ij)$ and thus the corresponding rotation vector to be parallel to the triad variable $B_{ij}$, as in the bulk. Therefore, we can interpret the corresponding rotation angle $\theta_{ij}$ as the (boundary) dihedral angle, and the action appearing in the amplitude as that of (causal) BF theory with an additional restriction on the relative direction of (boundary) connection and triad, characterizing a metric connection, or as the Regge action, just like in the bulk case. More precisely, our model gives naturally the transition amplitudes for (causal) BF theory or Regge action for fixed-B (metric) boundary conditions.\\

\ni Notice then that the measure factor is purely real in this case, which in turn has the consequence that there are \textit{no} quantum corrections to the bare Regge action appearing in the amplitude. This means that these corrections (which were discussed in \cite{newgfts}) are a consequence of our prescription for gluing simplices together, i.e. our choice of propagator, with the amplitude for each simplex being simply the bare Regge action. On the other hand, also the appearance of the Regge action itself is due, in our model,  to the choice of propagator, so that it does not look to be possible to have one without the other. The only possibility to do so would seem to be, maintaining our choice of kinetic term, a modification of the way one composes them with vertex functions and with each other, i.e. a modification of the Feynman rules or of Wick theorem. Such modification, however, look at present rather unnatural. It is also unclear, at the present stage, whether quantum corrections augmenting the bare action with higher order terms are a feature to be welcomed or avoided, but at least we are now able to identify more clearly their origin. \\

\ni Finally, the nice matching between  our amplitudes and those of (causal) BF theory is rather strongly dependent on the specific choice of observable (and thus of boundary data) we have made, leading to BF amplitudes with fixed-B boundary conditions. The choice of boundary conditions is specified by the boundary state/observable chosen. On the one hand, one could introduce a fixed non-trivial boundary connection by inserting a relative shift $R_{ij}\in SU(2)$ between the group elements associated to the edge $(ij)$ in the observable $O_{tet}(\phi)$. This group element will show up in the angle $\theta_{ij}(h,R)=\theta(G,R)$ in the action term, while the rotation $h_I h_i^{-1}$ will remain orthogonal to the edge variable $B_{ij}$. Such group element $R$ will therefore enter the action term in a similar way as the group element characterizing different choices of boundary conditions in usual BF theory (see \cite{EPR}). On the other hand, one could modify also the dependence of $O_{tet}$ on the Lie algebra variables $B$, for example averaging over them in order to reproduce the amplitude for BF with fixed connection on the boundary or with mixed boundary conditions. One realizes immediately, however, that in such generalized cases  the new symmetry of the GFT field does \textit{not} align `the $B$' variable associated to the edge with the corresponding boundary connection, as it appears in the action term.  and therefore we can not identify the action appearing in the amplitude above with the BF one. The reason for the mismatch between the discrete action which appears in the spinfoam amplitudes of the proposed GFTs and that of the discrete BF theory can be traced back to the kinetic term used in the GFT action. More specifically to the fact that we identify only the length of the $B$ variable with the `length' (the Laplacian) of the derivation on the group $J$. Such a kinetic term, therefore, seems to encode the correct dynamics of BF theory and of gravity, in the case of manifolds with boundaries, only for very special choices of boundary observables. Whether this is a correct feature of the GFT formalism or a true limitation of this specific class of GFT models remains to be understood. One could instead envisage a kinetic term where the vectors themselves are identified (as opposed to their lengths only), of the form $( B - J )^2$. Preliminary work on this `vectorial' kinetic term does seem to generate precisely the discrete BF action. Work on this will be reported elsewhere.\\

\ni A different, and promising, line of development, suggested by the mentioned limitations of the above model, but also making careful use of the insights it provides, in particular, the way the simplicial geometry is implemented at the level of the GFT action, and then in the amplitudes, will be mentioned in the concluding section.

\subsection{Lorentzian model}
\ni To construct a corresponding Lorentzian version of our GFT model, we only need to shift from $SU(2)$ to $SU(1,1)$ as our gauge group (and its Lie algebra). The definition of the model as well as the construction of the feynman amplitudes is essentially unchanged, and the same Feynman amplitudes maintain the same expression that identifies them as simplicial path integrals, now for a (causal) Lorentzian BF theory\footnote{The minor technical differences are a direct consequence of the fact that the Schroedinger kernel on SU(1,1) is basically the analytic continuation of the one on SU(2)}. The interested reader is invited to read the appropriate section \cite{newgfts} for details of the calculations. The only difference would be the new, `aligning' delta function coming from the different symmetry of the GFT field. The novel effect of this alignment between the $B$ and the $F$ is to kill the non-geometric (in the sense of not matching the relative spacetime characterization of $B$ and $F$ as spacelike or timelike vectors) configurations which were obtained in \cite{newgfts}. These were the configurations when the $B$ associated to an edge would be timelike while the holonomy around that edge would be a boost, similarly for a spacelike edge vs. rotational holonomy. Since in these situations the $B$ variable is not aligned with the holonomy they do not contribute to the path integral anymore. \\

\section{4d model}
\ni We close by discussing briefly the 4d analogue of the model presented above. This discussion is meant to consider the limitation of the generalised GFT formalism we presented from a different point of view. In fact, we show that, contrary to the 3d case, in 4d the model cannot reproduce 4d BF theory, while it may still have a chance to encode the dynamics of 4d gravity with a peculiar higher derivative action (due to quantum corrections), if the Plebanski constraint are correctly implemented on the B variables. This last point is discussed in detail in \cite{new4d}. With hindsight this could have been anticipated. The kinetic term of the new model, relaxing the conjugate nature of $B$ and $g$ variables already implies a different dynamics than that of BF theory, as testified by the quantum corrections to the BF action in the resulting simplicial path integral. This, however, may only imply that we are dealing with a BF theory with corrections to the bare action, which may not be a problem in itself. On the other hand, we have seen that the identification of the action appearing in the model, depending only on the modulus of the $B$ variables and the modulus of the holonomy associated to each dual face, with the BF action relied heavily on the geometric interpretation of the B variables themselves, on the corresponding interpretation of the restriction on the holonomy as a (partial) metricity condition, and of the holonomy itself as lying on the lane orthogonal to the $B$. In absence of simplicity constraints, no geometric interpretation of the $B$ variables is possible in the first place, and they do not identify a plane in $\mathbb{R}^4$, despite being associated to a triangle. Consequently, we could expect the restriction imposed by the new symmetry of the GFT field not to be sufficient, as a constraint on the holonomy, to match the action of the 4d model to the simplicial BF action (let alone the Regge action).\\

\ni The 4d model is obtained from the 3d one we presented by (see \cite{newgfts} for details; below, only the new features, coming from the restricted symmetry of the field, will be discussed): 

\begin{itemize}
\item Exchanging the (double cover of) the orthogonal/Lorentz group in three dimensions (SU(2) and SU(1,1) respectively) for the corresponding four dimensional analogues (Spin(4) $\simeq$ SU(2) $\times$ SU(2) and SL(2,$\mathbb{C}$) accordingly). The same applies of course for the Lie algebra. We will only discuss the Riemannian model below.
\item Taking the field to be a function of four pairs variables instead of three, thus the field of the theory is a function $\phi(g_1, \dots, g_4 ; B_1, \dots , B_4) : SU(2)^4 \times SU(2)^4 \times \mathfrak{su}(2)^4 \times \mathfrak{su}(2)^4 \rightarrow \mathbb{R}$. The field is now interpreted to represent a tetrahedron, with the group variables being interpreted as  being the parallel transports going from the center of the tetrahedron to the centres of the triangles, and the Lie algebra variables are thought of to be the discretized/`averaged' $B$ field (which is a 2-form now in 4d) over the corresponding triangles.
\end{itemize}

\ni The field is stipulated to be invariant under the analogue of (\ref{eq:invariance}), i.e. $\phi ( g_1 \, h^{-1} , g_2 \, h^{-1} , g_3 \, h^{-1}, g_4 \, h^{-1} ; h B_1 h^{-1}, h B_2 h^{-1}, h B_3 h^{-1} , h B_4 h^{-1}) = \phi (g_1 , g_2, g_3, g_4 ; B_1, B_2, B_3, B_4)$.
Finally, the action of the GFT is essentially the same as that in 3d, with the only difference being that the vertex has the combinatorics of a 4-simplex instead of a 3-simplex.

\ni Upon quantizaion, one obtains Feynman (fat) graphs which are 2-complexes locally dual to a 4-dimensional simplicial complex. The amplitudes again factorize per dual face and each dual face amplitude is of the form

\be
\label{eq:4damplitude}
A_{N_f} [ H_f , B_f ] = \mu ( [ \theta_f ] , | B_f | , N_f )  \, \delta(B_f - H_f B_f H_f^{-1})\,  e^{ i | B_f| | [  \theta_f]|}.
\ee

\ni Here $N_f$ is the number of dual vertices in the 2-face, $B_f$ is the $\mathfrak{spin}(4)$ element associated to the 2-face (or equivalently to the triangle dual to it), $H_f$ is the Spin(4) element giving the holonomy around the 2-face and $\theta_f = |\txt{log}(H_f)|$ (the length is computed with the Killing metric). The same convention regarding equivalence classes/branches of the logarithm applies here as well. Finally, $\mu(.)$ is a `measure' factor whose exact form will not concern us in what follows.\\

\ni In order to clarify the issue being discussed, we shall use the self-dual/anti-self-dual decomposition of Spin(4). Therefore $B = (B^+ , B^-)$ (we drop the label $f$ as we will only be concerned with a single dual face), where $B^+$ is the self-dual part of $B$ and $B^-$ is the anti-self-dual one. Of course each one of them can be naturally identified with an $\mathfrak{su}(2)$ element. Also, we will represent $H$ as $e^{F}$ with $F = (F^+, F^-)$ similarly to the case with the $B$'s, as well as $ H = (H^+ , H^-) = (e^{F^+} , e^{F^-} )$. From this it follows that

\bes
\nonumber
| B |^2 & = & (B^+)^2 + (B^-)^2, \;\;\;\;\;\;\;
| \, \theta \, | ^2  =  (F^+)^2 + (F^-)^2,\\
\label{eqnarray:separation}
\delta \Big [ B - H B H^{-1} \Big ] & = & \delta \Big [ B^+ - H^+ B^+ (H^+)^{-1} \Big ]  \, \, \, \delta \Big [ B^- - H^- B^- (H^-)^{-1} \Big ]
\ees

\ni Recalling the discussion in the 3d case, it follows immediately that the delta function (coming from the new, restricted symmetry of the field) does not quite align the $B$ field with the holonomy $H$ (what this means is that $B$ is not parallel, i.e. a multiple of, $F = \txt{log}(H)$). In fact, the general solution of the equation imposed by the delta function, seen as a condition on the holonomy, is: $H_f = e^{i \alpha \frac{B}{|B|} + i \beta *\frac{B}{|B|}} = e^{i \alpha \frac{B}{|B|}} e^{i\beta *\frac{B}{|B|}}$. Therefore, it only aligns $B^+$ with $F^+$ and $B^-$ with $F^-$. This has the important consequence that the discrete action appearing in the amplitudes (\ref{eq:4damplitude}) can \textit{not} be thought of as the BF action (not even a causal BF) since in the BF action one has the dot product of $B$ and $F$ while the obtained amplitudes have the product of the lenghts $|B| |F| \neq B \cdot F$ if $B \neq \lambda F$ (for some $\lambda$).\\

\ni A slightly different way of looking at the issue at hand is to note that in 3d after taking the new delta function into account, only one parameter is needed to parametrize the holonomy associated to an edge (in the bulk), which is naturally identified with the deficit angle. However, in 4d the new delta function does \textit{not} reduce the Spin(4) holonomy to a single parameter which one would want to identify with the deficit angle, not even in the bulk. Rather, the holonomy now generically depends on two angles, and so does the action, contrary to what happens in 1st order Regge calculus, where the only relevant angle is the deficit angle corresponding to the rotation in the plane orthogonal to the plane defined by the triangle. But the situation is also different from BF theory, whose action depends both on the total rotation angle of $H$ and on the relative direction of $B$ and $F$. \\

\ni Obviously, the first possibility that comes to mind to remedy to this situation is to impose an additional restriction on the fields forcing the $B$'s to be aligned with the corresponding $F$'s. However, there do not seem to be any obvious geometric/gauge invariances which one should take care of by incorporating them into the symmetries of the GFT field. Thus it is likely that the required restriction, even if it exists is rather artificial and ad hoc. Moreover, one could prove several no-go theorems which (mostly using gauge-invariance) show that the simplest ways of imposing such restriction are not allowed.\\

\ni Once more, this problematic form of the amplitudes, depending only on the modulus of $B$ and of $F$, comes directly from the GFT kinetic term. Therefore, one should investigate the model obtained by replacing the current kinetic term with the \lq\lq vectorial'' one mentioned above (i.e. the one whose corresponding classical equation equates the $B$ vector with the derivation on the group manifold).  Work on this modification of the GFT seems to give encouraging results and will be reported elsewhere.\\

\ni The other strategy  is to enforce the simplicity constraints in the model, to enforce with them the geometricity of the configurations summed over in the simplicial path integral. With this geometricity taken into account, the situation can be expected to be much closer to the 3d case and to the usual Regge calculus formulation of simplicial gravity. This turns out to be the case, with the caveat that both the $B$ variables and the group variables have to be constrained in the GFT action, possibly reflecting the fact that this generalised GFT formalism treats them on equal footing and as {\it independent} variables, thus relaxing their conjugate nature. The details of this work are reported in \cite{new4d}; here we will just summarize the main idea. As a way to impose the simplicity constraints, on top of constraining the $B$ variables on which each GFT field depends, at the level of the GFT action, one projects the field to the homogeneous space Spin(4)/SU(2) $\simeq S^3$ by averaging over the  diagonal SU(2) subgroup . The result is that one obtains the (product of) propagator(s) on the homogeneous space (the 3-sphere) and not on the full group, for each argument of the GFT field, and the dual face amplitude takes the form

\ben
\txt{Amplitude} \sim  \delta \Big [ B^+ - H^+ B^+ (H^+)^{-1} \Big ]  \, \, \, \delta \Big [ B^- - H^- B^- (H^-)^{-1} \Big ] \, \, \, e^{ i |B| \, | \vartheta(H)|},
\een

with addtional constraints on the $B$ variables being implicit here. In this formula $\vartheta$ is the `homogeneous' part of the holonomy $H$: $\vartheta (H ) = \txt{log} \Big [  H^+ (H^-)^{-1} \Big ]$.

\ni Now, keeping in mind the gauge invariance of the theory and the simplicity constraint, it is easy to see that there is a frame in which $B$ has the form
$B = (B^+ , B^- ) = \sqrt{2} |B| ( \hat{n} , - \hat{n})$,

where $\hat{n}$ is some unit vector in $\mathbb{R}^3$. The \lq\lq alignment delta function\rq\rq implies $F = (F^+ , F^-) = ( |F^+| \hat{n} , |F^-| \hat{n})$,
which in turn implies that $\vartheta( H ) = |F^+| - |F^-|$.
Putting all this together we see that the contribution $|B| \, | \vartheta(H)|$ of each dual face to the (bare) action is equal to 

\besn
|B| \,  \vartheta(H) & = &  \frac{1}{2 \sqrt{2}} \, \, B \cdot \, \Big (  |F^+| - |F^-| \Big ) \, ( \hat{n} , \hat{n} ) \,= \\
& = &  \frac{1}{2 \sqrt{2}} \, \, B \cdot ( |F^+ | \, \hat{n} , - |F^-| \, \hat{n} ) =  \frac{1}{2 \sqrt{2}} \,\, \, B\,\, \cdot \, \, F\,.
\eesn

\ni The last line, up to an irrelevant multiple, is just the BF action, with a constrained (by the simplicity constraints) $B$ variables. 
On the one hand, this result (presented in \cite{new4d}) shows that a much closer link between our Feynman amplitudes and simplicial BF theory and 4d gravity can be achieved. On the other hand, it confirms that the generalised GFT formalism we have presented here does not suffice to capture the properties of simplicial BF theory, before further modifications are imposed on its dynamics.

\section{Conclusions}
\ni We have shown that a new symmetry requirement on the GFT field, in the context of the extended GFT formalism presented in \cite{newgfts}, leads, in 3d, to Feynman amplitudes with a simplicial path integral form based on a simplicial action that can be nicely identified with the 3d Regge action in first order form, to a proper relation between the discrete connection and the triad vectors appearing in it, and to a much more satisfactory and transparent encoding of simplicial geometry already at the level of the GFT action. We believe that the geometric insights on the relation between simplicial connection and $B$ variables and on how this can be encoded in a GFT action will remain valid and useful in any further development in this area.\\

\ni We have also exposed the limitations of the same formalism: 1) the presence of additional contributions to the Regge action, whose origin we have clarified, but whose significance or necessity remains unclear; 2) a matching with 3d gravity transition amplitudes that seems to be obtainable only for a very restricted class of boundary observables/states; 3) the difficulty in obtaining Feynman amplitudes that could be related to BF theory (even with additional corrections to the action) in 4d. 
Most of these difficulties (just like most of the merits) found in this class of models seems to descend directly and inevitably from the choice of kinetic term we have adopted. Thus, remaining in the context of a generalized GFT formalism like this, with a GFT field depending on both group and Lie algebra elements, we have suggested that a different choice of kinetic term depending on all the components of the Lie algebra elements, and relating them to different derivative operators on the group manifold, could improve the situation and lead to Feynman amplitudes with a closer relation to simplicial BF theory. \\

\ni While all of the above concerns the improvement of our model within the same generalized GFT formalism, an altogether different line of development is also suggested by the above analysis. 
This new direction \cite{ioaristide1} stems from the idea that, instead of relaxing the conjugate nature (at the classical level) of $B$ and $g$ variables in the definition of the model, treating them on equal footing as arguments of the GFT field, as we do here, and as encoded in the kinetic term, one could instead take this conjugacy relation as the starting point for introducing the $B$ variables in the GFT formalism. This would mean to {\it map} the usual group field theories in which the field is a function of group elements into {\it non-commutative and non-local field theories} on several copies of the corresponding Lie algebra, using a generalized (non-commutative) Fourier transform, of the type developed in \cite{eteralaurent, eteralaurent1, laurent-majid, karim}. A first motivation for doing so is that, in the model we have presented, we are dealing with $\mathfrak{su}(2)$ as  an ordinary vector space, thus neglecting its non-commutative nature. This leads to suspect that, in doing so, we are neglecting some relevant information that should instead be incorporated in a correct model using the $B$ variables as arguments of the GFT field. More precisely, the non-commutative nature of the algebra would result in a non trivial star product for functions on it, when seen as functions on $\mathbb{R}^3$ (for $\mathfrak{su}(2)$). The use of a non-trivial star product for the products of fields in the GFT action, in turn, would result in a different composition rule for vertex amplitudes with propagators in the construction of the Feynman amplitudes of the model, and we have already noticed above how the rather puzzling quantum corrections $S_c$ to the BF action are the result of the way our individual vertex (3-simplex) amplitudes (itself given by a simple BF action)  compose.\footnote{This non-commutative Fourier transform has been recently applied to the construction of 4d spinfoams \cite{eteravalentin}.} Finally, we mention another hint for the need to move to a non-commutative setting, arising from the analysis of the model we have presented. It has recently been shown, both in 3d and in 4d \cite{emergent, emergent1}, that effective scalar field theories on flat non-commutative spaces emerge naturally from group field theories, with the group manifold underlying the GFT providing the momentum degrees of freedom of the effective matter field, while the non-commutative position variables are expected to be related to the conjugate $B$ variables. A similar analysis performed using the generalized GFT formalism we have presented in this paper shows \cite{ioalessandro} that, while everything works fine even for these generalised GFT models as long as one focuses on the group sector, the expected matter field theory fails to emerge if one focuses instead on the Lie algebra sector of the same GFT models. This failure can be indeed traced back to having neglected the non-commutative structure of the same Lie algebra at the level of the fundamental GFT, i.e. in the definition of the generalised formalism itself, and to the use of the kinetic term relaxing the conjugate nature of $B$ and $g$ variables. This suggests again the non-commutative representation developed in \cite{ioaristide1}.

\section*{Acknowledgements}
We thank A. Baratin and V. Bonzom, for discussions and comments on this work. 
The work of DO is supported by a Sofja Kovalevskaja Prize from the Alexander von Humboldt Foundation, which is gratefully acknowledged.

\end{document}